\documentclass[journal]{IEEEtran}
\usepackage{authblk}

\usepackage{amsmath}
\usepackage{hyperref}
\usepackage{braket}
\usepackage{graphicx}
\usepackage{microtype}

\ifCLASSINFOpdf
\else
\fi

\begin{document}

\title{Diffusive Description of Superconducting Single-Photon Detectors Using Quantum Mechanics}

\author[1]{Krzysztof Pomorski }
\author[2]{Eryk Halubek}

\affil[1]{University of New Mexico, Center for High Technology Materials, Albuquerque, New Mexico, USA } 
\affil[2]{Warsaw University of Technology, Faculty of Physics, Warsaw, Poland}

\maketitle

\begin{abstract}
The methodology of non-Hermitian quantum mechanics is employed to parameterize the dynamics of coupled classical diffusion equations describing low- and high-energy quasiparticles in superconducting single-photon detectors in one to three spatial dimensions. 
\end{abstract}


\IEEEpeerreviewmaketitle

\section{Quasiparticle dynamics in SNSPDs and their mapping onto quantum mechanics}

Transition-edge detectors used as superconducting single-photon or neutron detectors \cite{Schilling1}, \cite{SCNeutron} operate on the principle that an incident event, such as a photon or neutron striking a current-carrying superconductor, induces a localized transition from the superconducting state to the normal state. The deposited energy creates a resistive hot spot within the superconducting material and produces high- and low-energy quasiparticles (QPs), whose concentrations are denoted by $C_e$ and $C_{qp}$, respectively. These quasiparticles obey Fermi--Dirac statistics. As the bias current continues to flow, the formation of the resistive region gives rise to a measurable voltage across the device. The associated Joule heating can further enlarge the hot spot. Subsequently, thermal relaxation in the cryogenic environment allows the detector to return to its superconducting state. As the hot spot shrinks and eventually disappears, the voltage signal decays and nondissipative current flow is restored. We first formulate the classical equations of motion and then map them onto a dissipative quantum system.

To model this process, we use the coupled linear diffusion equations introduced in \cite{Schilling1}. These equations can be formulated using the Laplacian operator in one, two, or three spatial dimensions (and extended to higher dimensions): $\nabla_1^2=\frac{d^2}{dx^2}$, $\nabla_2^2=\frac{d^2}{dx^2}+\frac{d^2}{dy^2}$, and $\nabla_3^2=\frac{d^2}{dx^2}+\frac{d^2}{dy^2}+\frac{d^2}{dz^2}$. The resulting system of coupled partial differential equations is
\begin{eqnarray}
\frac{d}{dt}
\begin{pmatrix}
C_{e}(\vec{x},t) \\
C_{qp}(\vec{x},t)
\end{pmatrix}
=
\begin{pmatrix}
D_e \nabla^2 & 0 \\
B(t) & [ D_{qp} \nabla^2-A ]
\end{pmatrix}
\begin{pmatrix}
C_{e}(\vec{x},t) \\
C_{qp}(\vec{x},t)
\end{pmatrix}
\end{eqnarray}
Here, $B(t)=\frac{ \xi h v }{ \Delta \tau_{qp} }e^{-\frac{t}{\tau_{qp}}}$ and $A=\frac{1}{\tau_r}$, where $\Delta$ is the superconducting order parameter, $\tau_r$ and $\tau_{qp}$ are characteristic quasiparticle lifetimes, $v$ is the frequency of the incident photon, and $\xi$ denotes the detector efficiency. The preceding equation shows that the dynamics of the high-energy particles, described by $C_e$ (subsystem A), are independent of the dynamics of the low-energy quasiparticles, described by $C_{qp}$ (subsystem B). In contrast, the dynamics of subsystem B depend on those of subsystem A (system $A$ $\rightarrow$ system $B$). Multiplying both sides of the coupled equations by the imaginary unit $i=\sqrt{-1}$ yields a representation in terms of dissipative quantum dynamics:
\begin{eqnarray*}
i\frac{d}{dt}
\begin{pmatrix}
C_{e}(\vec{x},t) \\
C_{qp}(\vec{x},t)
\end{pmatrix}
= i
\begin{pmatrix}
D_e \nabla^2 & 0 \\
B(t) & [ D_{qp} \nabla^2-A ]
\end{pmatrix}
\begin{pmatrix}
C_{e}(\vec{x},t) \\
C_{qp}(\vec{x},t)
\end{pmatrix}, \nonumber \\
i\frac{d}{dt}
\ket{\psi}_{q}
= i
\begin{pmatrix}
D_e \nabla^2 & 0 \\
B(t) & [ D_{qp} \nabla^2-A ]
\end{pmatrix}
\ket{\psi}_{q},
\end{eqnarray*}
%
%

This provides a time-dependent, non-Hermitian description of dissipative quantum evolution and leads to the following discretized computational scheme:
\begin{eqnarray*}
i\frac{d}{dt}
\begin{pmatrix}
C_{e}(\vec{x},t) \\
C_{qp}(\vec{x},t)
\end{pmatrix}
= \hat{H}(t)
\begin{pmatrix}
C_{e}(\vec{x},t) \\
C_{qp}(\vec{x},t)
\end{pmatrix},
\nonumber \\
\hat{H}(t)=
-i
\begin{pmatrix}
-D_e \frac{d^2}{dx^2} & 0 \\
-B(t) & [ -D_{qp} \frac{d^2}{dx^2} +A ]
\end{pmatrix}, \nonumber \\
\begin{pmatrix}
C_{e}(x,t) \\
C_{qp}(x,t)
\end{pmatrix}=
\ket{\psi(t)} = e^{\frac{1}{i} \int_{t0}^{t}dt_1 \hat{H}(t_1)  } \ket{\psi(t_0)} = \nonumber \\
=Exp \Bigg[ \int_{t0}^{t}du  
\begin{pmatrix}
D_e \frac{d^2}{dx^2} & 0 \\
B(u) & [ D_{qp} \frac{d^2}{dx^2} -A ]
\end{pmatrix}
\Bigg]
\begin{pmatrix}
C_{e}(x,t_0) \\
C_{qp}(x,t_0)
\end{pmatrix} \approx \nonumber \\ 
\approx \Pi_{r=1}^{r=s-1}
e^{ \Bigg[ \int_{t_{r}}^{t_{r+1}}du  
\begin{pmatrix}
D_e \frac{d^2}{dx^2} & 0 \\
B(u) & [ D_{qp} \frac{d^2}{dx^2} -A ]
\end{pmatrix}
\Bigg] }
\begin{pmatrix}
C_{e}(\ x,t_0) \\
C_{qp}( x,t_0)
\end{pmatrix} \approx \nonumber \\
 \Pi_{r=1}^{r=s-1}
e^{ \Bigg[  \frac{t-t_0}{s-1}  
\begin{pmatrix}
D_e \frac{d^2}{dx^2} & 0 \\
B(t_0+\frac{ (r-1)(t-t_0)}{s-1}) & [ D_{qp} \frac{d^2}{dx^2} -A ]
\end{pmatrix}
\Bigg]}
\begin{pmatrix}
C_{e}(x,t_0) \\
C_{qp}(x,t_0)
\end{pmatrix}, 
\end{eqnarray*}
Here, $r$ indexes the time subdivisions between $t_0$ and $t$. Over each interval of duration $\Delta t = \frac{t-t_0}{s-1}$, the Hamiltonian is treated as approximately constant, which simplifies the time integration.  

\section{Methodology for the spectral analysis of a dissipative Hamiltonian}

Consider a time-dependent dissipative Hamiltonian with $N_{max}$ complex-valued, time-dependent eigenvalues and corresponding eigenvectors satisfying the orthonormality condition $\braket{\psi_{E_{i,k}}(t)|\psi_{E_{i,s}}(t)}=\delta_{k,s}$. Under this assumption, we use the following spectral representation:
\small
\begin{eqnarray}
\hat{H}(t) 
= \sum_{i_1=1}^{N_{max}}\sum_{i_2=1}^{N_{max}}E(i_1 \rightarrow i_2)(t)\ket{\psi_{E_{i_2}}(t)}\bra{\psi_{E_{i_1}}(t)} .
\end{eqnarray}

\normalsize


For a finite time increment $\Delta t$, transitions between energy eigenstates, including unchanged states, are characterized by
$E(i_1 \rightarrow i_2)(t,\Delta t)=\bra{ \psi_{E_{i_1}}(t+\Delta t)} \hat{H}(t) \ket{ \psi_{E_{i_2}}(t) }$ .
The linearity of the coupled diffusion equations motivates a superposition of states even when an SNSPD absorbs several photons simultaneously or nearly simultaneously. We therefore write the final state as 

\begin{equation}
\ket{\psi}= ( a_1\ket{\psi_{E1}} + a_2 \ket{\psi_{E2}} + a_3 \ket{\psi_{E3}} + .. ),
\end{equation}
where $a_1, \ldots, a_s$ are chosen to ensure nonnegative low- and high-energy QP concentrations. 

\section{Quantum particle in a box and classical coupled diffusion equations}
\begin{figure}
\includegraphics[scale=0.2]{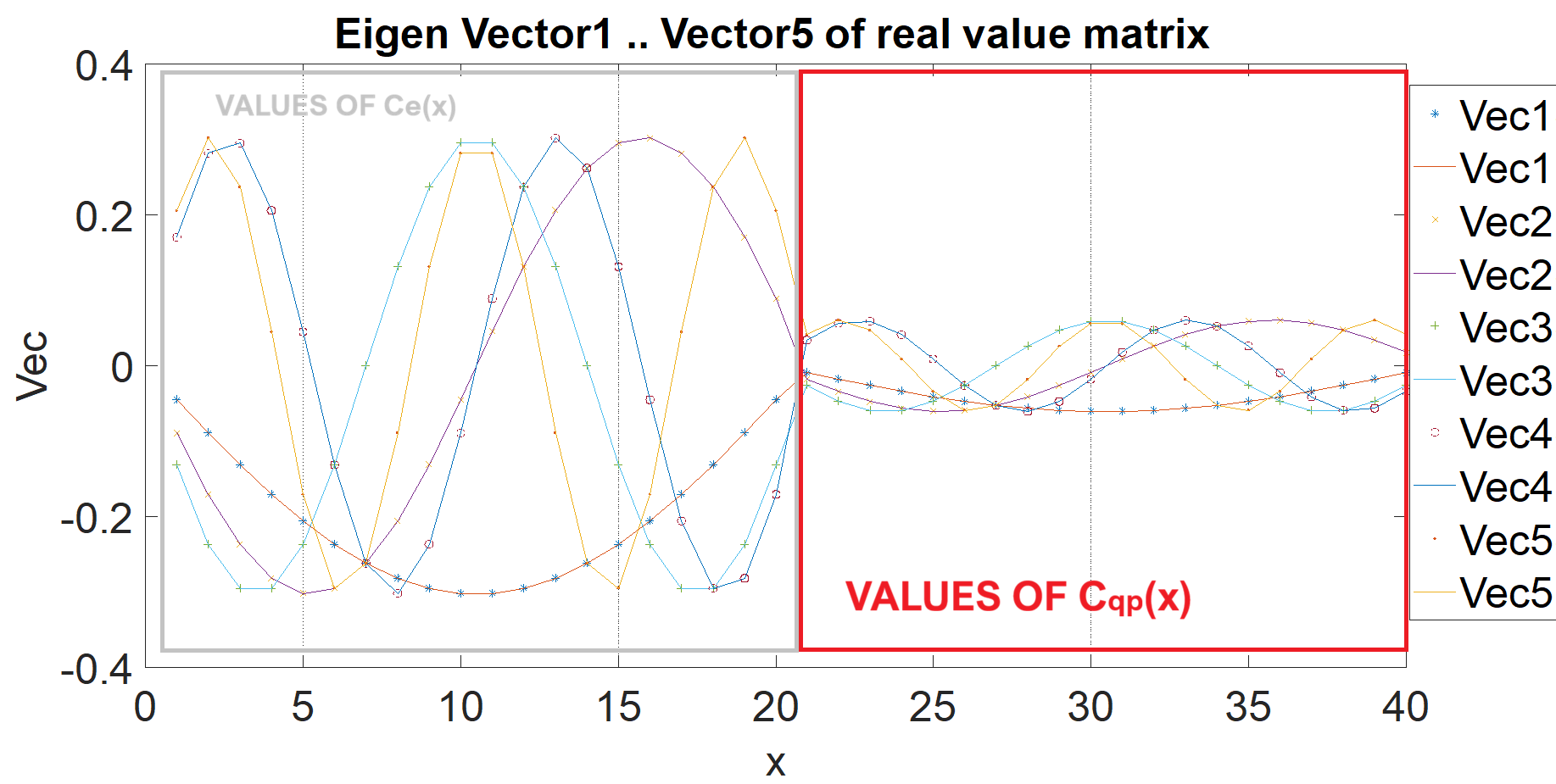} 
\caption{Eigenvectors of the linearly coupled diffusion equations corresponding to the first three energy eigenvalues and resembling the wavefunctions of a particle in a box with infinitely high potential walls. The left half of the horizontal axis corresponds to the distribution of $C_e$, whereas the right half corresponds to that of $C_{qp}$.}\label{fig:eigenvectors}
\end{figure}

The Schr\"odinger particle-in-a-box model with infinitely high potential walls can be related to the classical coupled diffusion equations. We write 
\begin{eqnarray*}
\hat{H}(t)=
-i
\begin{pmatrix}
[ -D_e \frac{d^2}{dx^2} ]_{ \text{ qpa in box} }  & 0 \\
-B(t) & [ -D_{qp} \frac{d^2}{dx^2} +A ]_{ \text{qpa in box} }
\end{pmatrix}, \nonumber \\
\end{eqnarray*}
Consequently, the corresponding energy eigenvalues are purely imaginary and can be expressed as follows:  
\begin{eqnarray*}
i\frac{d}{dt}C_{e}(x,t) = -i [ -D_e \frac{d^2}{dx^2} ] C_{e}(x,t) = 
\nonumber \\ 
( -i E_0) \ket{E_0}\bra{E_0} + (-i E_1) \ket{E_1}\bra{E_1} + (-i E_2 )\ket{E_2}\bra{E_2} + ..,
\end{eqnarray*}
For a particle confined to $x \in [0, L]$, the eigenfunctions are $\psi_{Es}(x)=\sqrt{ \frac{2}{L} }  sin( \frac{ n \pi x }{L})= \sqrt{ \frac{2}{L}}  \frac{1}{2i} ( e^{i (\frac{ n \pi x }{L}) } - e^{-i (\frac{ n \pi x }{L}) } )$, with the corresponding energy eigenvalues
 $E_s= \frac{s^2 \pi^2 \hbar^2}{2mL^2}$. 
The transformation $E_k \rightarrow -i E_k$ makes the real-valued energies purely imaginary while preserving the eigenfunction profiles (Fig.~\ref{fig:eigenvectors}). The left and right halves of the horizontal axis correspond to $C_e$ and $C_{qp}$, respectively. The coupled system can thus be interpreted as a particle in a square-well potential evolving in reversed imaginary time. The Gaussian distribution discussed in \cite{Schilling1} then takes the form 

\begin{eqnarray}
e^{\frac{-i E_0 (t-t_0)}{i }}a_0 \psi_{E0}(x,t) + e^{\frac{-i E_1 (t-t_0)}{i }}a_1 \psi_{E1}(x,t_0)+ .. = \nonumber \\
=e^{- E_0 (t-t_0)}a_0 \psi_{E0}(x,t) + e^{- E_1 (t-t_0)}a_1 \psi_{E1}(x,t_0)+ .. = \nonumber \\
= \frac{2}{\sqrt{D t} ( \text{erf}\left(\frac{L}{2 D t}\right))} \frac{N_0}{ \sqrt{ 4 \pi D t} }{e^{-\frac{(x-\frac{L}{2})^2}{D t}}}u(x,t) = \ket{C_e(x,t)}.
\end{eqnarray} 
The Hamiltonian of the coupled subsystems A and B satisfies 
\begin{eqnarray}
\begin{pmatrix}
\hat{D}_1 & 0 \\
-B(t) & \hat{D_2}
\end{pmatrix}=i \hat{H}.
\end{eqnarray}
The eigenvalues $\lambda_1$ and $\lambda_2$ of the diagonal blocks $\hat{D}_1$ and $\hat{D}_2$ are
\begin{eqnarray}
\lambda_1=Eigenvalue[\hat{D_1}]=-Eigenvalue[-D_e(\frac{d^2}{dx^2}+\frac{d^2}{dy^2})], \nonumber \\
\lambda_2=Eigenvalue[\hat{D_2}]=-Eigenvalue[-D_{qp}(\frac{d^2}{dx^2}+\frac{d^2}{dy^2}+A)].
\end{eqnarray}
The state norm of the coupled subsystems A and B is not generally conserved. We write the state as 
\begin{eqnarray}
\begin{pmatrix}
C_{e}(\vec{x},t) \\
C_{qp}(\vec{x},t)
\end{pmatrix}
=
\begin{pmatrix}
1 \\
\frac{B(t)}{(\lambda_1-\lambda_2)}
\end{pmatrix}
e^{\int_{t0}^{t} [D_e \nabla^2]dt_1 }C_{e}(\vec{x},t_0) = \nonumber \\ 
= 
\begin{pmatrix}
1 \\
\frac{\frac{ \xi h v }{ \Delta \tau_{qp} }e^{-\frac{t}{\tau_{qp}}}}{(\lambda_1-\lambda_2)}
\end{pmatrix}
e^{\int_{t0}^{t} [D_e \nabla^2]dt_1 }C_{e}(\vec{x},t_0) 
= \ket{\psi(\vec{x},t)} .
\end{eqnarray}
Because $C_{qp}$ may decay faster than $C_e$, the latter can remain nonzero when $C_{qp}$ is nearly zero, as observed numerically in one to three spatial dimensions. The energy spectrum of $C_e(\vec{x},t)$ uniquely determines that of $C_{qp}(\vec{x},t)$. Distinct spatial distributions can arise for a nonuniform superconducting order parameter ($\Delta(x) \neq \mathrm{const}$), as described by time-dependent Ginzburg--Landau theory. 

\section{Conclusions and future directions}
We have established a correspondence between classical coupled diffusion equations and dissipative quantum systems for superconducting single-photon detectors. We also propose a numerical representation of the dissipative quantum dynamics in one to three spatial dimensions using products of matrix exponentials. 
\section{Acknowledgment}
We acknowledge financial support from the NQVL:QSTD:Pilot project, ``National Virtual Laboratory for Photonic Quantum Computing'' (NSF Award No.~2410813).

\ifCLASSOPTIONcaptionsoff
  \newpage
\fi

\end{document}